\documentclass[sn-mathphys-num]{sn-jnl}

\usepackage{graphicx}%
\usepackage{multirow}%
\usepackage{amsmath,amssymb,amsfonts}%
\usepackage{amsthm}%
\usepackage{mathrsfs}%
\usepackage[title]{appendix}%
\usepackage{xcolor}%
\usepackage{textcomp}%
\usepackage{manyfoot}%
\usepackage{booktabs}%
\usepackage{algorithm}%
\usepackage{algorithmicx}%
\usepackage{algpseudocode}%
\usepackage{listings}%

\begin{document}

\title{Open problems in information geometry: a discussion at FDIG 2025}
 \author*[1]{Tomonari Sei}
 \author[2]{Hiroshi Matsuzoe}
 \affil[1]{
 Department of Mathematical Informatics,
 Graduate School of Information Science and Technology, 
 The University of Tokyo}
 \affil[2]{
Department of Computer Science,
Graduate School of Engineering,
 Nagoya Institute of Technology}

\abstract{Open problems in information geometry are collected and discussed in the conference ``Further Developments of Information Geometry (FDIG) 2025'' held at the University of Tokyo, Japan, from March 18 to 21, 2025.}

\maketitle


\section{Jun-ichi INOGUCHI (Hokkaido University) (communicated by Hiroshi MATSUZOE)}
The following problems were provided by Professor Jun-ichi Inoguchi at Hokkaido University and interted by Hiroshi Matsuzoe of the Organizing Committee.
\begin{enumerate}
\item Determine the moduli space of left invariant affine connections 
on the statistical Lie group of univariate normal distributions
compatible to the Fisher metric. 
(Comment by Matsuzoe: This problem was independently considered by Kobayashi, et al.\cite{Kobayashi2025})
\item Characterize the exponential connection and the mixture connection of 
the statistical manifold of univariate normal distributions by variational problems.
\item Let $\mathcal{N}$ the statistical 
Lie group of multivariate normal distributions \cite{Kobayashi_Ohno}. 
Are there any left invariant semi-Riemannian metrics on 
$\mathcal{N}$ whose Levi-Civita connection is 
the exponential connection.  Note that the answer is no for 
the statistical Lie group of univariate normal distributions 
(see \cite{Inoguchi24}). 
\item 
The statistical manifold of binomial distributions is regarded as the 
standard example of $1$-dimensional Hessian manifold (see \cite{Inoguchi25}). Investigate $1$-dimensional Hessian manifolds, especially 
$1$-dimensional Hessian manifolds derived from probability distribution families.
\item 
Let $(M,g,\nabla)$ be a statistical manifold 
and 
$(N,\nabla^{\prime})$ be an affine manifold. 
A smooth map $\psi:M\to N$ is said to be 
a \emph{harmonic map} if its 
\emph{tension field} $\tau(\psi)=\mathrm{tr}_{g}(\nabla d\psi)$ vanishes. 
Here $\nabla d\psi$ is defined by
\[
(\nabla d\psi)(X,Y)=\nabla^{\prime}_{d\psi(X)}d\psi(Y)-d\psi(\nabla_{X}Y)  
\]
for all vector fields $X$ and $Y$ on $M$. 
In Riemannian geometry, 
the domain manifold is chosen as a 
Riemannian manifold $(M,g,\nabla^g)$ equipped with 
the Levi-Civita connection $\nabla^g$ and the target manifold is a 
Riemannian manifold $(N,h,\nabla^h)$ equipped with 
Levi-Civita connection $\nabla^h$. 
Under this assumption, $\psi$ is 
harmonic if and only if 
it is a critical 
point of the Dirichlet energy over any compact region $D$ of $M$. 

According to \cite{Inoguchi-Ohno}, a smooth map $\psi:M\to M$ 
from a statistical manifold $(M,g,\nabla)$ into itself is said to be a 
\emph{statistically harmonic map} if 
it is a harmonic map from $(M,g,\nabla)$ into 
$(M,g,\nabla^{*})$, where $\nabla^{*}$ is the conjugate 
connection of $\nabla$ with respect to $g$.

In Riemannian geometry, 
the identity map $\mathrm{id}$ is always harmonic. 
The stability of identity maps have been studied 
differential geometers. In particular, 
the stability of identity maps of compact semi-simple Riemannian symmetric spaces 
has similarity to the Yang-Mills stability 
over compact semi-simple Riemannian symmetric spaces. 
In the case of statistical manifolds, $\mathrm{id}$ is not always harmonic. 
Indeed, the tension field of 
the identity map of a statistical manifold 
is $\tau(\mathrm{id})=\mathrm{tr}_{g}K$, where $K=\nabla-\nabla^g$ is 
the difference tensor field. Thus the so-called \emph{apolarity} 
of a statistical manifold is characterized as the statistical harmonicity 
of the identity map. Based on these observations, 
we propose here 
\begin{itemize}
\item Investigate statistical manifolds with parallel $\tau(\mathrm{id})$.
\item Investigate statistical manifolds whose identity map is a critical 
point of the bienergy functional. For a smooth map $\psi:(M,g,\nabla)\to 
(M,g,\nabla^*)$, its bienergy over a relatively 
compact region $D$ of $M$ is 
\[
E_{2}(\psi;D)=\int_{D}
\frac{1}{2}|\!|\tau(\psi)|\!|^2\,\mathrm{d}v_g.
\]
\end{itemize}

\end{enumerate}

\section{Sosuke Ito (The University of Tokyo)}
As a study of nonequilibrium thermodynamics, we have now investigated the relations between the $1$- or $2$-Wasserstein distance and the entropy production rate for the Markov jump process described by the master equation (or the entropy production rate for the chemical reaction described by the rate equation). Since the Benamou-Brenier formula~\cite{benamou2000computational} for the Fokker-Planck equation is closely related to the problem of the minimum entropy production rate in nonequilibrium thermodynamics~\cite{ito2024geometric}, we need to consider several generalizations of the Benamou-Brenier formula and related results for the master equation or the rate equation (see Refs.~\cite{yoshimura2023housekeeping,kolchinsky2024generalized,nagayama2024infinite}).

We now imagine that some generalizations of the Benamou-Brenier formula including the definition by Jan Maas~\cite{maas2011gradient}, might be possible by considering the generalized mean, such as the Stolarsky mean, as we did in Ref.~\cite{nagayama2024infinite}. Interestingly, the Legendre duality can be introduced for any generalized mean (see Appendix E in Ref.~\cite{nagayama2024infinite}). The logarithmic mean seems to be special because the dual dissipation functions are both quadratic. This question may also be related to the variety of the large deviation theory~\cite{kaiser2018canonical}, which uses the Legendre duality for the geometric mean. So I wonder whether the treatment in information geometry is helpful or not for the relationship between the generalized mean and the variety of the optimal transport theory. The possible questions are as follows.
\begin{itemize}
    \item To consider the optimal transport theory for the Markov jump process~\cite{maas2011gradient}, Jan Maas introduced the logarithmic mean between two probabilities or two flows. The logarithmic mean seems to be natural when considering the gradient flow or the heat equation for the Markov jump processes. Can we understand this special mathematical property of the logarithmic mean as an analog of the Chentsov's theorem for the Fisher metric, or the special property of the Kullback-Leibler divergence in information geometry? Can we think of the dual structure for the generalized mean as an analog of information geometry? Are information-geometric treatments such as Amari-Chentsov connections useful for understanding the variants of the optimal transport theory for the Markov jump processes with a unified perspective?
\end{itemize}

Such an interesting mathematical structure by the generalized mean can also be seen in quantum information theory.
A possible related question is as follows.
\begin{itemize}
    \item In quantum information theory, the variety of the quantum Fisher metric is introduced by using the generalized mean~\cite{petz2011introduction}. Is there a framework for treating this variety in information geometry?
\end{itemize}
\section{Tetsuya J. KOBAYASHI (The University of Tokyo)}

In conventional information geometry for static statistical problems and others, we have at least three constructions of the dually flat structure:
\begin{enumerate}
\item Differential geometric construction starting from a manifold equipped with a Riemannian metric $g$ and a connection $\nabla$\cite{amari2000,Frank2020}.
\item Divergence based construction starting from a manifold and a contrast function\cite{Eguchi1983,Eguchi1997}
\item Contact geometric construction\cite{Goto2016,Nakajima2021}
\end{enumerate}
When we extend information geometry from static problems to dynamics, the knowledge from other fields working on diffusion processes, optimal transport, Markov jump processes, and chemical reaction networks suggest that another metric or dually flat structure is required on the dual pair of tangent and cotangent spaces for velocity and force driving the dynamics defined on the base manifold equipped with conventional (another) dually flat structure\cite{Kobayashi2024}.

What would be the purely information geometric and mathematically sound construction for such a joint-structure for dynamics?

\section{Takeru MATSUDA (The University of Tokyo)}

Csisz{\'a}r \cite{Csiszar1975} is one of the early results of {information geometry}.
He showed that the Sinkhorn algorithm for the matrix scaling problem \citep{Idel2016} can be interpreted as alternating e-projections and minimizes the Kullback--Leibler divergence.
Recently, a quantum (non-commutative) generalization of {matrix scaling} called {operator scaling} has been found to appear in surprisingly many fields of mathematics, physics and computer science \citep{Idel2016}.
Gurvits \cite{Gurvits2004} extended the {Sinkhorn algorithm} to operator scaling.

Since the Sinkhorn algorithm for matrix scaling is equivalent to alternating minimization of the Kullback--Leibler divergence, it is natural to ask if some quantum analogue of the Kullback--Leibler divergence yields the same characterization of the operator Sinkhorn algorithm.
Recently, \cite{MS2022} investigated the operator Sinkhorn algorithm from the viewpoint of {quantum information geometry} and showed that the operator Sinkhorn algorithm is interpreted as alternating e-projections with respect to the {symmetric logarithmic derivative (SLD) metric}. 
However, due to the non-vanishing torsion in the space of density matrices, it is still open whether the operator Sinkhorn algorithm can be written as alternating minimization of some divergence.

\section{Hiroshi MATSUZOE (Nagoya Institute of Technology)}

\begin{enumerate}
\item Elucidate what geometric structures are essential to information geometry. 

For example, let $(M,g)$ be a Riemannian manifold, $\mathcal{U} = (U; x^1, \dots, x^n)$ and $\mathcal{V} = (V; y^1, \dots, y^n)$ local coordinate systems on $M$.  We say that $\mathcal{U}$ and $\mathcal{V}$ are \emph{bi-orthogonal} if
\begin{equation} \label{biorthogonal}
  g\left(\frac{\partial}{\partial x^i}, \frac{\partial}{\partial y^j}\right) = \delta_{ij} 
\end{equation}
for ${}^{\forall}i, {}^{\forall}j =1, \dots, n$ on $U\cap V$. It is well-known that, if $(M, g, \nabla, \nabla^*)$ is a dually flat space, then 
there exist bi-orthogonal coordinates such that $\mathcal{U}$ and 
$\mathcal{V}$ are dual affine coordinates \cite[Chapter 3]{amari2000}.
However, since \eqref{biorthogonal} is given only in local coordinate systems and a Riemannian metric, there is no need to assume the existence of affine connections.
Give a necessary and sufficient condition for bi-coordinates to exist on a Riemannian manifold 
(cf. \cite[\S 19]{Ch-book}). 
How can we state the generalized Pythagorean theorem if we do not assume a pair of mutually dual affine connections (cf. \cite[\S 22]{Ch-book}, \cite{Csiszar1975})?
\item Let $(M,g)$ be a Riemannian manifold and $\nabla$ an affine connection on $M$. The \emph{dual connection} (or \emph{conjugate connection}) $\nabla^*$ of $\nabla$ is defined by
\begin{equation} \label{dual}
    Xg(Y, Z) = g(\nabla_XY,Z) + g(Y,\nabla^*_XZ),
\end{equation}
where $X, Y$ and $Z$ are arbitrary vector fileds on $M$.

Conversely, suppose that $\nabla$ and $\nabla^*$ are affine connections on $M$. What is a sufficient condition for two affine connections to be dual to each other? 
Namely, for two given affine connections $\nabla$ and $\nabla^*$, can we find a Riemannian metric $g$ which satisfies Equation \eqref{dual}?
\item Give a good definition of the "divergence'' function in information geometry.

Many divergence functions have been studied in information geometry, such as KL-divergence and Bregman divergence.
The geometry of dual connections does not require the positivity of the divergence, and the geometry of singular statistical models does not require the fact that the Riemannian metric is induced from the second derivative of the divergence.
What should the divergence function be in information geometry?
The notion of canonical divergence has been studied by many authors. See \cite{HK2000, FA2021}, for example.

\item Give good examples of the use of $\alpha$-connections where $\alpha$ is not $\pm 1$.

Let $S$ be a statistical model, $g^F$ the Fisher metric on $S$, and $\nabla^{(\alpha)}$ the $\alpha$-connection on $S$. We call $(S, g^F, \nabla^{(\alpha)})$ an
\emph{invariant statistical manifold}.
If $\nabla^{(\alpha)}$ is flat, we say that a dually flat space $(S, g^F, \nabla^{(\alpha)}, \nabla^{(-\alpha)})$ is \emph{invariant}.  
If $S$ is an exponential family, $(S,g^F, \nabla^{(1)}, \nabla^{(-1)})$ is an invariant dually flat space.

Suppose that $S_q = \{p_q(\ast;\mu,\sigma)\} \ (q\in(1,3))$ is the
set of $q$-Gaussian distributions
(the set of Student's $t$-distributions with degree of freedom $\nu = (3-q)/(q-1)$), where 
$\mu \in (-\infty, \infty)$ is a location paramter, 
and $\sigma \in (0,\infty)$ is a scale paramter.
Set $A := (2q-1)/(2-q)$. Then $(S_q, g^F, \nabla^{(A)}, \nabla^{(-A)})$ is an invariant dually flat space
with $A \ne 1$ \cite{mori2020, matsuzoe2025}.

Are there any useful examples of invariant 
dually flat spaces
with $\alpha \ne \pm 1$?

Note that the KL-divergence always induces $(g^F, \nabla^{(1)}, \nabla^{(-1)})$. Therefore, the KL-divergence dose not induce
a dually flat structure on $S_q$.
\end{enumerate}

\subsection{Comment by Shiro Ikeda (The Institute of Statistical Mathematics)}

Finding good examples of the use of $\alpha \ne \pm 1$, as discussed in point $4$, is both interesting and important for the study of information geometry.

When we consider a set of exponential family ($e$-flat, $\alpha = 1$) distributions $\{p_{\theta}\}$, the corresponding dual connection ($\alpha = -1$) is essential in statistics. Assuming the data are i.i.d. samples from an unknown distribution, the empirical distribution $\hat{p}$ can be formed, and the maximum likelihood estimation corresponds to the $m$-projection ($-1$-projection) from $\hat{p}$ to the submanifold $\{p_{\theta}\}$. Here, the i.i.d. assumption, especially the ``independence,'' is crucial for deriving the MLE because the product form of $p_{\theta}$ becomes simple.

Instead of exponential family distributions, let us consider the set of $q$-Gaussian distributions, which are $A$-flat with $A = (2q-1)/(2-q)$, and an associated estimation problem. As an extension of the $\alpha = \pm 1$ case, we may consider the corresponding ``$-A$''-projection. However, the i.i.d. assumption does not lead to this projection. 

The main logic of modern statistics is based on the i.i.d., or at least the independence assumption, and the asymptotic theory. If we try to find any good examples of the use of $\alpha \ne \pm 1$ connections, any such example must be founded not on independence, but on some alternative principle. I am not aware of any example in statistics. There might be some related topics in statistical physics.
\section{Hiroshi NAGAOKA (The University of Electro-Communications)}
\begin{enumerate}
\item
Mathematically rigorous treatment of the geometry for the totalities of probability distributions on infinite sample spaces satisfying certain conditions is an important subject for infinite-dimensional extension of IG, and since the seminal work of Pistone and his group, a variety of studies have been made on the subject. 
Most of these studies, however, seem to be based on purely mathematical/geometrical interests, with little interest in applications of (or relations to) probabilistic/statistical problems. 
(Please forgive me if I have missed any significant results in this regard.)  
Below, I would like to present two problems in probability theory/statistics that are, I think, worthwhile to shed light on from the viewpoint of infinite-dimensional geometry. 
\begin{enumerate}
\item 
In the case of finite sample spaces, Sanov's theorem and its relation (equivalence in some sense) to Cram\'{e}r's theorem 
via the Legendre transformation are particularly important in that they show that the dually flat IG structure underlies a problem of fundamental importance in probability theory.  It is known (see Chap.~6 of  \cite{dembo_zeitouni} and the references cited there) that 
Sanov's theorem and Cramer's theorem have been extended to the case when the sample space is an arbitrary Polish space.  There, the expression of the relative entropy (the KL divergence) in terms of the Legendre transformation is derived by a rigorous functional-analytic argument.  Then a natural question arises as to whether these results are related to some infinite-dimensional \emph{differential geometric} structure. 
\item 
Finite-dimensional statistical models on infinite sample spaces are the most popular objects in traditional statistics.
It may be worth investigating whether treating such models as submanifolds of the infinite-dimensional manifold can help us understand or simplify statistical discussions about them.  For instance, in the traditional asymptotic theory of parameter estimation, we often assume several complicated regularity conditions on models. Is it possible to express such conditions as  some  geometric conditions on submanifolds?
\end{enumerate}
Note: One of the most important fact that we have realized through our attempts to extend IG to the quantum setting is 
that ``parameter estimation" and ``large deviation/hypothesis testing" 
are essentially different kinds of problems, and that the reason why a single IG structure,  the dually flat structure consisting of the Fisher metric and e, m-connections,  is applicable to both of them in the classical case should not be found in the relationship between the two problems, but should be seen as a fortunate coincidence due to C\v{e}ncov's uniqueness theorem. This viewpoint may be also important in studying the infinite-dimensional extension of IG in connection with probabilistic/statistical applications. 
\item Let $\mathcal{S} = \mathcal{S}  (\mathcal{H} )$ 
be the totality of strictly positive density operators on a finite-dimensional Hilbert space $\mathcal{H}$, and let $g, \nabla^{(\rm{m})}$ and $\nabla^{(\rm{e})}$ be a monotone Riemannian metric on $\mathcal{S}$, the flat connection on $\mathcal{S}$ induced by the natural affine embedding $\mathcal{S} \hookrightarrow \mathcal{L}_h$, 
where $\mathcal{L}_h$ denotes the totality of Hermitian operators on $\mathcal{H}$, and 
the dual connection of $\nabla^{(\rm{m})}$ with respect to $g$, respectively. 
The problem presented here is to determine the $\nabla^{(\rm{e})}$-autoparallel submanifolds of $\mathcal{S}$ or, equivalently,  to find the condition for a pair $(\rho, V)$ of a point $\rho\in \mathcal{S}$ and 
a linear subspace $V$ of the tangent space $T_\rho ( \mathcal{S})$ to have a  $\nabla^{(\rm{e})}$-autoparallel submanifold which contains $\rho$ with the tangent space $V$. 
The problem can be regarded as seeking  (at least formal) quantum analogues of exponential families, but 
the nonvanishing torsion of $\nabla^{(\rm{e})}$ makes the problem highly nontrivial. 
A more general form of the problem is to determine the $\nabla$-autoparallel submanifolds of a 
Riemannian manifold with a connection $\nabla$ whose dual connetion is flat. 
We know very little about the problem at present except for a few special or obvious cases 
(see \cite{nagaoka_fujiwara_autoparallel}). 
\end{enumerate}

\section{Tomonari SEI (The University of Tokyo)}

Efron \cite{Efron_1978} found an example of an exponential family in which the space of expectation parameters is not convex.
Are there systematic studies on this phenomenon? If not, can we develop?

In the discussion of FDIG 2025, Dr.~T.-K.~L.~Wong pointed out that the problem is related to non-regular exponential families \cite{BN1978}.
The potential function $\kappa$ of an exponential family $P_\theta(dx)=e^{\theta\cdot x-\kappa(\theta)}P_0(dx)$ is called steep if the gradient map $\nabla\kappa$ diverges at every boundary point of ${\rm dom}(\kappa)$.
Define the expectation parameter space by $\mathfrak{T}=\{\nabla\kappa(\theta)\mid \theta\in{\rm int}({\rm dom}(\kappa))\}$.
In Theorem 9.2 of \cite{BN1978}, it is shown that $\kappa$ is steep if and only if $\mathfrak{T}$ coincides with the interior of the convex closure of the support of $P_0$.
Hence, non-convex expectation parameter spaces appear only if $\kappa$ is non-steep;
see also Section~9.8 (i) of \cite{BN1978}.
In particular, the support of $P_0$ has to be unbounded.

\begin{figure}[htbp]
\centering
\includegraphics[width=0.5\textwidth]{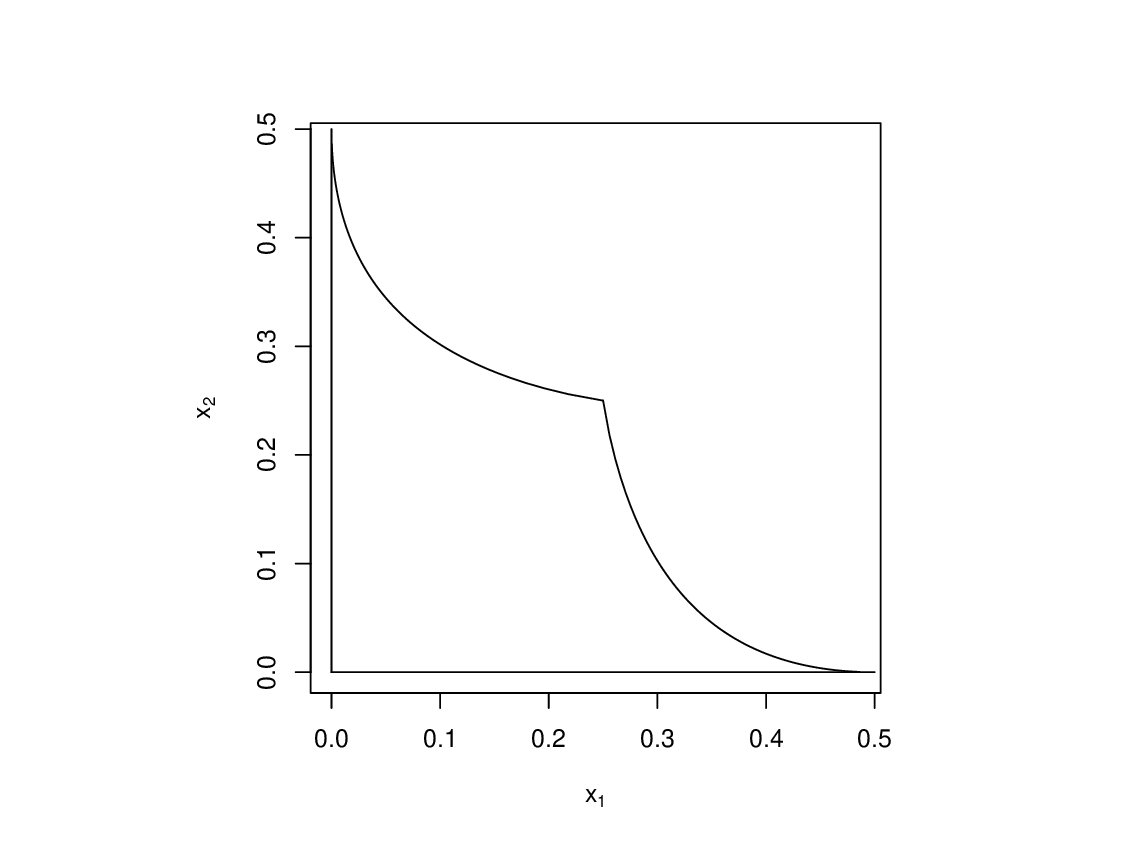}
\caption{An example of a non-convex and bounded expectation parameter space.}
\label{fig:sei_1}
\end{figure}

Here, we give an example of non-convex and bounded $\mathfrak{T}$. Consider
\[
  P_0(dx) = \frac{1}{2}\frac{3}{(1+x_1)^4}\nu_1(dx) + \frac{1}{2}\frac{3}{(1+x_2)^4}\nu_2(dx),
\]
where $\nu_1$ and $\nu_2$ are the Lebesgue measure on $\mathbb{R}_+\times \{0\}$ and $\{0\}\times\mathbb{R}_+$, respectively.
The domain of $\kappa$ is $(-\infty,0]^2$.
Then, $\mathfrak{T}$ is non-convex and bounded as shown in Figure~\ref{fig:sei_1}.

How to characterize possible shapes of non-convex expectation parameter spaces?

\section{Ting-Kam Leonard WONG (University of Toronto)}
\begin{enumerate}
\item The topic of generalized exponential families has received a lot of attention in information geometry. A key example is the $q$-exponential family \cite{amari2011geometry} which was reparameterized as a $\lambda$-exponential family (where $q = 1 - \lambda$) in \cite{wong2022tsallis}:
\begin{equation} \label{eqn:lambda.exp.family}
p_{\lambda}(x; \theta) = (1 + \lambda \theta \cdot F(x))^{1/\lambda} e^{-\varphi_{\lambda}(\theta)}.
\end{equation}
In \cite{wong2022tsallis}, it was shown under suitable regularity conditions that a $\lambda$-exponential family, say $M_{\lambda}$, can be equipped with a natural dualistic geometry which has constant sectional curvature $\lambda$. Similarly, in the literature a generalized exponential family is usually studied under the assumption that the deformation parameter (which is nonparametric for e.g.~the $\phi$-exponential family) is fixed. Is there a systematic way to study the deformation parameter as another parameter? In the context of \eqref{eqn:lambda.exp.family}, we have a foliated statistical manifold $M := \bigcup_{\lambda} M_{\lambda}$, where each leaf $M_{\lambda}$ is ``well-behaved''. Can we equip $M$ with a natural geometric structure that is useful for statistical inference? Here we note the recent work \cite{uohashi2022extended} in which a divergence was constructed on a  foliation by deformed probability simplexes.
\item It is well-known that the squared $2$-Wasserstein distance between two Gaussian distributions $\mu = N(a, A)$ and $\nu = N(b, B)$ is given by $\mathcal{W}_2^2(\mu, \nu) = |a - b|^2 + d_{\mathrm{BW}}^2(A, B)$, where $d_{\mathrm{BW}}^2(A, B) = \mathrm{tr}(A) + \mathrm{tr}(B) - 2 \mathrm{tr}\left( A^{\frac{1}{2}}BA^{\frac{1}{2}} \right)$ is the squared Bures-Wasserstein distance. The Wasserstein geometry of Gaussian distributions was studied in \cite{T10}. In \cite{gelbrich1990formula}, Gelbrich proved that if $\mu$ and $\nu$ are probability measures on $\mathbb{R}^n$ with finite second moments, then $\mathcal{W}_2(\mu, \nu) \geq \mathcal{W}_2(\tilde{\mu}, \tilde{\nu})$, where $\tilde{\mu}$ is the Gaussian distribution whose first and second moments match those of $\mu$, and $\tilde{\nu}$ is defined similarly. Gelbrich's lower bound has found many applications including distributionally robust optimization (see e.g.~\cite{kuhn2019wasserstein}). Now, suppose that we restrict $\mu$ and $\nu$ to be elliptical distributions. A non-degenerate elliptical distribution $\mu$ (with finite second moment) on $\mathbb{R}^n$ can be parameterized by a mean vector $m$, a covariance matrix $\Sigma$, and a radial distribution $\nu$ on $\mathbb{R}_+$ with $\int_{\mathbb{R}_+} r^2 d \nu(r) = n$ (we write $\mu = \mathcal{E}_2(m, \Sigma, \nu)$). If we let $X = m + R \Lambda S$, where $S$ is spherically distributed, $\Lambda \Lambda^{\top} = \Sigma$, $R \sim \nu$ and is independent of $S$, then $X \sim \mathcal{E}_2(m, \Sigma, \nu)$. We may regard the space of all elliptical distributions as a semiparametric statistical manifold. Although there are no closed form formulas for the $2$-Wasserstein distance between two elliptical distributions $\mu_0 = \mathcal{E}_2(m_0, \Sigma_0, \nu_0)$ and $\mu_1 = \mathcal{E}_2(m_1, \Sigma_1, \nu_1)$ (with different radial distributions), it is not difficult to show the upper bound
\[
\mathcal{W}_2^2(\mu_0, \mu_1) \leq |m_0 - m_1|^2 + d_{\mathrm{BW}}^2(\Sigma_0, \Sigma_1) + \frac{1}{n} \mathrm{tr}\left( \left( \Sigma_0^{\frac{1}{2}} \Sigma_1 \Sigma_0^{\frac{1}{2}} \right)^{\frac{1}{2}}\right) \mathcal{W}_2^2(\nu_0, \nu_1).
\]
Can we develop an improved Gelbrich lower bound that involves also $\nu_0$ and $\nu_1$? Furthermore, can we study the Wasserstein geometry of elliptical distributions in the spirit of \cite{T10} (also see \cite{amari2024information})?
\end{enumerate}
\bibliography{reference}

\end{document}